\documentstyle[aps,12pt]{revtex}
\begin{document}
\draft
\author{O. B. Zaslavskii}
\address{Department of Physics, Kharkov Karazin's National University, Svobody Sq.4,\\
Kharkov\\
61077, Ukraine\\
E-mail: aptm@kharkov.ua}
\title{{\bf Nonextreme black holes near the extreme state and} {\bf acceleration
horizons: thermodynamics and quantum-corrected geometry }}
\maketitle

\begin{abstract}
We consider the class of metrics that can be obtained from those of
nonextreme black holes by limiting transitions to the extreme state such
that the near-horizon geometry expands into a whole manifold. These metrics
include, in particular, the Rindler and Bertotti - Robinson spacetimes. The
general formula for the entropy of massless radiation valid either for
black-hole or for acceleration horizons is derived. It is argued that, as a
black hole horizon in the limit under consideration turns into an
acceleration one, the thermodynamic entropy $S_{q}$ of quantum radiation is
due to the Unruh effect entirely and $S_{q}=0$ exactly. The contribution to
the quasilocal energy from a given curved spacetime is equal to zero and the
only nonvanishing term stems from a reference metric. In the variation
procedure necessary for the derivation of the general first law, the metric
on a horizon surface changes along with the boundary one, and the account
for gravitational and matter stresses is an essential ingredient of the
first law. This law confirms the property $S_{q}=0$. The quantum-corrected
geometry of the Bertotti - Robinson spacetime is found and it is argued that
backreaction of quantum fields mimics the effect of the cosmological
constant $\Lambda _{eff\text{ }}$ and can drastically change the character
of spacetime depending on the sign of $\Lambda _{eff}$ --- for instance,
turn $AdS_{2}\times S_{2}$ into $dS_{2}\times S_{2}$ or $Rindler_{2}\times
S_{2}$. Two latter solutions can be thought of as the quantum versions of
the cold and ultracold limits of the Reissner-Nordstrom-de Sitter metric.
\end{abstract}

\pacs{PACS numbers: 0470B, 0470D, 9760L, 0420J, 0440N }

\section{introduction}

In recent years, distinction between extreme and nonextreme black holes
became an object of intensive studies. A nontrivial peculiarity of the
extreme state consists in that its properties depend crucially on the way
the limiting transition is performed. Consider, for instance, the
Reissner-Nordstr\"{o}m black hole as the simplest example. If its mass $m$
is put equal to its charge $e$, the proper distance between the horizon and
any other point outside it diverges, the topology corresponding to the
annulus of an infinite size. As a result, the vicinity of the horizon
responsible for the entropy does not contribute to the Euclidean action.
This property gave grounds to reason that classical extreme black holes
possess unusual thermodynamic properties in the sense that their temperature
is arbitrary and is not connected with the Hawking one (which is zero in the
limit in question), while the entropy $S=0$ \cite{hawking}. On the other
hand, as was shown in \cite{zasl1}, \cite{zasl2}, there exists a limiting
transition $m\rightarrow e$ in the topological sector of nonextreme black
holes such that the proper distance between the horizon and points outside
it remains finite. Although either the surface gravity or the Hawking
temperature in this case are equal to zero, the physical temperature
defining properties of a thermodynamic ensemble is finite, and the entropy $%
S $ has the Bekenstein-Hawking value $A/4$, where $A$ the a horizon area.

Thus, we have two types of limiting states of black holes with essentially
different properties. For the sake of shortness we will further refer to the
first type as 1 and to the second type as 2. Types 1 and 2 possess different
Euler characteristics \cite{bin}. A sharp distinction between them manifests
itself not only in thermodynamics and Euclidean approach but also in
Lorentzian quantum geometrodynamics \cite{claus}. In fact, the states of
type 2 represent the direct product of two-dimensional spaces --- for
example, $AdS_{2}xS_{2}$ (see below in a more detail). In this sense, they
are not real black holes. However, as they arise as a result of the limiting
transition from the true ones whose horizons certainly possess thermodynamic
properties, it is necessary to elucidate what happens to these properties in
the limiting states.

Up to now our treatment concerned classical black holes. Meanwhile, studying
quantum effects in the background of extreme black holes is of special
interest. If a black hole is in state 1 there are good grounds to believe
that the principal conclusion $S=0$ made in \cite{hawking} for classical
black holes loses its validity when effects of backreaction from quantum
fields surrounding a hole are taken into account. These effects force the
temperature to take the Hawking value $T_{H}=0$ since otherwise the
stress-energy tensor of quantum fields diverges on a horizon \cite{anders},
so the possibility of thermodynamic description of extreme black holes
becomes questionable. In this respect, we are faced with the curious
situation when even weak switching on dynamical interaction between a hole
and its quantum environment changes drastically thermal properties of the
system.

Correspondingly, the question arises about the role of quantum effects for
state 2. In the paper \cite{ent} for the particular case of the Bertotti -
Robinson (BR) metric \cite{br}, which is the limiting form of the
Reissner-Nordstr\"{o}m black hole \cite{zasl2}, the quantum correction to
thermodynamic entropy $S_{q}$ was found. It turned out that this correction
does not change the thermodynamic properties of a system dressed by a
quantum field as compared with a bare one and, moreover, $S_{q}=0$. In the
present paper we generalize this result and show that it retains its
validity for a wide set of the limiting state belonging to class 2
independently of the particular form of the metric. Thus, the role of
quantum effects in thermal properties proves to be very different for both
types of states.

We consider also the influence of quantum effects on geometrical properties
of the type 2 states. First of all, even in the absence of quantum effects
the nature of a horizon in state 2 is different from that of a black hole
whose limiting form the state 2 represents. Thus, for the BR metric, the
existence and properties of a horizon is an observer-dependent effect due to
geodesic incompleteness in some accelerated frames \cite{lap}, so the
horizon of a Reissner-Nordstr\"{o}m black hole turns into an acceleration
one. The geometry of quantum-corrected acceleration horizons possesses two
main features. It retains the general form of the direct product of two
two-dimensional spacetimes typical of the classical counterpart and, in this
sense, our approach confirms the recent observation made for the particular
case of the geometry which represents the direct product of anti-de$%
%TCIMACRO{\func{Si}}
%BeginExpansion
\mathop{\rm Si}%
%EndExpansion
$tter space with a sphere \cite{ads}. However, we argue that the concrete
type of geometry may be changed by quantum backreaction in an essential way.
For example, it can lead to the appearance of a second, cosmological-like
horizon which is absent for a classical counterpart of the metric.

In turn, the general structure of a metric and, in particular, the
difference between an acceleration horizon as compared to a black hole one,
affects the thermodynamics in what concerns the formulation of the general
first law. In the black hole case, the usual picture implies that the
horizon radius is a free parameter allowed to vary, whereas the metric on a
boundary of a system may be kept fixed. However, for a typical metric with
an acceleration horizon the radius of a sphere entering an angular part of a
metric is constant, so if it is kept fixed the first law turns into an empty
identity. Therefore, we are faced here with a somewhat unusual situation
when the first law acquires a nontrivial meaning only under condition that a
boundary metric itself is varied. Another peculiarity consists in that the
quasilocal energy density of such a system coming from a gravitational
action is identically zero and nonzero contribution stems entirely from a
reference metric (usually chosen as a flat one).

The paper is organized as follows. In Sec. II we derive the general
expression for the entropy of quantum massless radiation valid in spacetimes
with either black hole horizons or acceleration ones and show that in the
latter case, $S_{q}=0$. This derivation relies on the definition of the
stress-energy tensor, its conservation law, the scale properties of massless
radiation, and does not use the first law.

In Sec. III we give an qualitative explanation to the found property $%
S_{q}=0 $ as connected with the Unruh effect in, generally speaking, curved
spacetimes. We discuss relationship between different pairs of spacetimes
which represent Minkowski - Rindler analogues in curved manifolds and show
how in some limit this analogy becomes literal coincidence.

In Sec. IV, we show that if in the formulation of the first law one properly
accounts for spatial gravitational stresses and the pressure of quantum
fields, this law confirms the property $S_{q}=0$.

In Sec. V, we derive the explicit form of the quantum-corrected geometry and
find that there are three qualitatively different types of solutions
depending on whether the curvature of submanifold in time-radial directions
is negative, positive or zero.

In Sec. VI, we summarize briefly the results obtained and mark some possible
problems for future research.

\section{entropy of hawking radiation in terms of stress-energy tensor}

Consider the Euclidean metric of the form 
\begin{equation}
ds^{2}=d\tau ^{2}b^{2}+\alpha ^{2}dy^{2}+r^{2}(y)d\omega ^{2}\text{.}
\label{genmet}
\end{equation}
If $r$ is not constant, it can be chosen as a new radial variable and we
arrive at a generic spherically symmetrical spacetime. The special class of
metric arises when $r=const\equiv r_{+}$. It just corresponds to the
limiting transition to the extreme state of nonextreme black holes since in
that limit all points of a manifold take the same value of $r$ \cite{zasl2}.
In the particular case $b=r_{+}\sinh y$, $\alpha =r_{+}$ we obtain the BR
spacetime. While deriving in this section the formula for quantum entropy,
we will not specify further the coefficients $b$ and $\alpha $.

The total entropy of the system is equal to 
\begin{equation}
S=S_{0}+S_{q}\text{.}  \label{total}
\end{equation}
Here $S_{0}$ is the Bekenstein-Hawking entropy $S_{0}=A/4$ where $A$ is the
area of the event horizon and $S_{q}$ is the entropy of Hawking radiation.
While $S_{0}$ appears as the result of the zero-loop approximation in the
path-integral approach \cite{gib}, $S_{q}$ is due to quantum fields.The
quantity $S_{0}$ is determined solely by one characteristics of the event
horizon and in this sense manifests the universality of laws of black hole
physics. On the contrary, the quantity $S_{q}$ depends strongly on the kind
of fields and concrete details of matter distribution. If $r$ is not
constant, the expression for the entropy of Hawking radiation in terms of
the renormalized stress-energy tensor is obtained for a wide class of metric 
\begin{equation}
ds^{2}=d\tau ^{2}f(\frac{r_{+}}{r})+dr^{2}f^{-1}+r^{2}d\omega ^{2}\text{, }%
d\omega ^{2}=d\theta ^{2}+\sin ^{2}\theta d\phi ^{2}  \label{met}
\end{equation}
in \cite{rad}. Here $r_{+}$ is a horizon radius. The Schwarzschild metric
belongs to this class with $f(x)=1-x$. In the state of thermal equilibrium
the entropy of massless quantum field in this background inside a cavity of
a radius $r_{B}$ is equal to \cite{rad} 
\begin{equation}
S_{q}=16\pi ^{2}r_{+}\left| f^{\prime }(1)\right|
\int_{+}^{r_{B}}drr^{2}(T_{r}^{r}-T_{0}^{0}-T_{\mu }^{\mu }\ln \frac{r_{B}}{r%
})\text{,}  \label{ent1}
\end{equation}
where $T_{\mu }^{\nu }$ are the components of the renormalized stress-energy
tensor in the Hartle-Hawking state. As such a tensor is finite on a horizon
in an orthonormal frame, all components entering (\ref{ent1}) are finite and
the entropy converges. It is worth stressing that this is just thermodynamic
entropy which reveals itself in physical experiments but not a
statistical-mechanical part of it which for black holes has no direct
physical meaning and even diverges \cite{frolov}. This expression contains,
apart from two first terms typical of a classical thermal gas, also a purely
quantum anomaly part which is necessary for the general first law to hold 
\cite{rad}.

Meanwhile, we are willing to obtain the entropy of quantum field valid for
the BR spacetime 
\begin{equation}
ds^{2}=r_{+}^{2}(d\tau ^{2}\sinh ^{2}x+dx^{2}+d\omega ^{2})  \label{ber}
\end{equation}
and its generalization (\ref{genmet}). As the coefficient at $d\omega ^{2}$
may now be constant, we cannot use the result (\ref{ent1}) of \cite{rad}
directly. Below we suggest derivation suitable for both cases, when $r$ can
be either variable or constant.

It is convenient to normalize the radial coordinate in such a way that $y=0$
at a horizon and $y=1$ on the boundary, the period of the Euclidean time $%
\beta _{0}$ is chosen to be $2\pi $. It is implied that a system is situated
in a cavity of a finite size. Let the surface area of a horizon be equal to $%
\pi r_{+}^{2}$. We choose $y=l/l_{B}$, $l$ is a proper distance from a
horizon $(l=l_{B}$ for a boundary) and assume that metric coefficients take
the form 
\begin{equation}
r=r_{+}q(l/r_{+})\equiv r_{+}q(zy)\text{, }b=r_{+}d(l/r_{+})\equiv r_{+}d(zy)%
\text{, }\alpha =l_{B}\text{, }z=l_{B}/r_{+}  \label{coef}
\end{equation}
which embraces both cases (\ref{ber}) and (\ref{met}). Now we will show how
the expression for the entropy can be recovered from the components of the
stress-energy tensor.

Let us consider the variation of the metric which preserves the form (\ref
{coef}), so only parameters of the metric are allowed to change. This metric
has two such parameters - for instance, $z$ and $r_{+}$. Then the variation
of the Euclidean action $I$ of quantum field inside a cavity splits to two
parts: $\delta I=\delta _{1}I+\delta _{2}I$. Here the first term has the
standard form: 
\begin{equation}
\delta _{1}I=\frac{1}{2}\int d^{4}x\sqrt{g}T_{\mu \nu }\delta g^{\mu \nu }%
\text{,}  \label{1}
\end{equation}
$x^{0}=\tau $, $x^{1}=y$, $x^{2}=\theta $, $x^{3}=\phi $. It is worth
bearing in mind, however, that this term does not exhausts the total
variation since the formula (\ref{1}) implies that the metric on a boundary
is fixed. Meanwhile, we consider generic variation which affects the
boundary surface. Below we will see that the term $\delta _{2}I\sim \delta
r_{B}$ can be recovered from the scale properties of the action.

The terms with the variation of the metric can be written as 
\begin{equation}
T_{\mu \nu }\delta g^{\mu \nu }=-2T_{0}^{0}\frac{\delta b}{b}%
-(T_{2}^{2}+T_{3}^{3})\frac{\delta r}{r}-T_{1}^{1}\frac{\delta \alpha }{%
\alpha }\text{.}  \label{var}
\end{equation}
It follows from (\ref{coef}) that $\delta b/b=\delta r_{+}/r_{+}+\delta
zyd^{\prime }/d$, $\delta r/r=\delta r_{+}/r_{+}+\delta zyq^{\prime }/q$, $%
\delta \alpha /\alpha =\delta z/z+\delta r_{+}/r_{+}$, where the prime
denotes the derivative with respect to argument. Performing integration over
angle variables and Euclidean time we obtain for the first part of
variation: $\delta _{1}I/8\pi ^{2}=-\frac{\delta r_{+}}{r_{+}}\int_{0}^{1}dy%
\sqrt{\tilde{g}}T_{\mu }^{\mu }-\delta z\int_{0}^{1}dyy\sqrt{\tilde{g}}%
[T_{0}^{0}\frac{d^{\prime }}{d}+T_{1}^{1}z^{-1}+(T_{2}^{2}+T_{3}^{3})\frac{%
q^{\prime }}{q}]$ where $\sqrt{\tilde{g}}=b\alpha r^{2}=r_{+}^{4}zq^{2}d$
takes into account the fact that the factor $8\pi ^{2}$ due to integration
over angles and Euclidean time is already singled out. Now let us make use
of the conservation law $T_{1;\nu }^{\nu }\equiv \frac{(T_{1}^{\mu }\sqrt{g})%
}{\sqrt{g}}-\frac{1}{2}\frac{\partial g_{\alpha \beta }}{\partial x_{1}}%
T^{\alpha \beta }=0$. Taking into account the explicit form of the metric (%
\ref{genmet}) and scale properties of it we have: $\frac{(q^{2}dT_{1}^{1})^{%
\prime }}{q^{2}d}-[\frac{d^{\prime }}{d}T_{0}^{0}+(T_{2}^{2}+T_{3}^{3})\frac{%
q^{\prime }}{q}]=0$. By substituting this expression into $\delta _{1}I$ we
obtain after integration by parts: 
\begin{equation}
\frac{\delta _{1}I}{8\pi ^{2}}=-\frac{\delta r_{+}}{r_{+}}\int_{0}^{1}dy%
\sqrt{\tilde{g}}T_{\mu }^{\mu }-\delta z\frac{\sqrt{\tilde{g}}}{z}%
T_{1}^{1}(z)\text{.}  \label{var1}
\end{equation}
Let us write down also the term $\delta _{2}I/8\pi ^{2}\equiv \gamma \delta
r_{B}$. It follows from (\ref{coef}) that $\delta r_{B}=\delta
r_{+}q+q^{\prime }r_{+}\delta z=\delta r_{+}\frac{r_{B}}{r_{+}}+q^{\prime
}r_{+}\delta z$, whence 
\begin{equation}
\delta I/8\pi ^{2}=\delta r_{+}(\gamma q-\frac{1}{r_{+}}\int_{0}^{1}dy\sqrt{%
\tilde{g}}T_{\mu }^{\mu })+\delta z[\gamma q^{\prime }r_{+}-\frac{\sqrt{%
\tilde{g}}}{z}T_{1}^{1}(z)]\text{.}  \label{del}
\end{equation}
This general formula describes the response of the action to the change of
two variables $r_{+}$ and $z$.

Now let us take into account scale properties of the action of the massless
quantum field. It is a dimensionless quantity which must depend on
dimensionless arguments. In general, one can compose two such combinations
from the parameters of the problem: $z$ and $l_{0}/r_{+}$, where $l_{0}$ is
the Planck length. Correspondingly, one can write down $I=I(z,l_{0}/r_{+})$.
However, in the semiclassical approximation, when $l_{0}\ll r_{+}$ and all
effects of high-order loops are neglected, only the first parameter is
relevant: $I=I(z,0)\equiv I(z)$ plus negligible corrections. (Let me recall
that in the Scwharzschild case when either the entropy or the action of
quantum fields are calculated explicitly, they depend on one variable $%
r_{B}/r_{+}$ only, where $r_{B}$ is a radius of a system \cite{mat};
meanwhile, now we use $z=l_{B}/r_{+}$ instead of $r_{B}/r_{+}$). Thus, the
semiclassical action of massless fields depends only on one variable $z$: $%
I=I(z)$. I\ stress that this fact, as we will see below, is consistent with
the presence of quantum anomaly terms in the action.

This means that the coefficient at $\delta r_{+}$ should be equal to zero,
whence 
\begin{equation}
\gamma =(r_{+}q)^{-1}\int_{0}^{1}dy\sqrt{\tilde{g}}T_{\mu }^{\mu }\text{.}
\label{gamma}
\end{equation}
By substitution into the part proportional to $\delta z$ we obtain 
\begin{equation}
\delta I/8\pi ^{2}=\delta z[\frac{q^{\prime }}{q}\int_{0}^{1}dy\sqrt{\tilde{g%
}}T_{\mu }^{\mu }-\frac{\sqrt{\tilde{g}}}{z}T_{1}^{1}]\text{.}  \label{2del}
\end{equation}
From dimension grounds it also follows that the stress-energy tensor has the
general form 
\begin{equation}
T_{\mu }^{\nu }=r_{+}^{-4}t_{\mu }^{\nu }(r/r_{+})\equiv r_{+}^{-4}f_{\mu
}^{\nu }\text{,}  \label{tensor}
\end{equation}
where $t_{\mu }^{\nu }(q(zy))\equiv $ $f_{\mu }^{\nu }(zy)$. In eq. (\ref
{2del}) the value of $T_{1}^{1}$ is to be taken at the boundary where $y=1$.
After simple transformations we have 
\begin{equation}
\frac{1}{8\pi ^{2}}\frac{dI}{dz}=\frac{q^{\prime }}{q}%
\int_{0}^{z}dxd(x)q^{2}(x)f_{\mu }^{\mu }(x)-f_{1}^{1}(z)d(z)q^{2}(z)\text{.}
\label{der}
\end{equation}
Now we may find $I$ by direct integration. The constant of integration is
determined by the demand that $I=0$ when $z=0=l_{B}$ (no room for
radiation). Changing the order of integration in the first term we may get
rid of a double integral. Using the thermodynamic formula $I=-\int d^{4}x%
\sqrt{g}T_{0}^{0}-S_{q}$ let us write down the result at once for the
entropy: 
\begin{equation}
S_{q}=\int_{0}^{z}dxq^{2}(x)d(x)[f_{1}^{1}-f_{0}^{0}-f_{\mu }^{\mu }\ln 
\frac{q(z)}{q(x)}]\text{.}  \label{S}
\end{equation}
Returning to dimensional variables we may rewrite this formula as 
\begin{equation}
S_{q}=\int dV_{4}(T_{1}^{1}-T_{0}^{0}-T_{\mu }^{\mu }\ln \frac{r_{B}}{r})
\label{2S}
\end{equation}
Here $dV_{4}=\beta _{0}d^{3}x\sqrt{g}$ is the element of Euclidean
four-volume which in our particular coordinate system has the form (after
integration over angles and time variable) $dV_{4}$ $=8\pi
^{2}dxr_{+}^{3}q^{2}(x)d(x)$ where $x=yz$. Let me stress that the
contribution of conformal anomalies, as seen from (\ref{2S}), is taken into
account.

The formula (\ref{2S}) holds for any spacetime whose metric has the form (%
\ref{coef}). There are two typical classes of it. In the first case, the
metric can be rewritten as 
\begin{equation}
ds^{2}=d\tau ^{2}U(r)+dr^{2}V^{-1}(r)+r^{2}d\omega ^{2}\text{.}
\label{spher}
\end{equation}
where $U$ and $V$ depend on $r_{+}$ via combination $r_{+}/r$. In
particular, the Schwarzschild metric belongs to this class, in which case $%
\beta _{0}=4\pi r_{+}$, $dV_{4}=16\pi ^{2}r_{+}r^{2}dr$ and we return to the
result (\ref{ent1}) describing a nonextreme black hole. The second class can
be obtained from (\ref{spher}) by a well-defined limiting transition to the
extreme state such that a local Tolman temperature remains finite nonzero
quantity at any point outside a horizon. In so doing, $r\rightarrow r_{+}$
for all points of manifolds. As a result, the metric takes the form (\ref
{genmet}) with $r(y)=r_{+}=const$ and $b$ typically representing a
combination of hyperbolic functions (see \cite{zasl2} for details). In so
doing, the components $T_{\mu }^{\nu }$ of the stress-energy tensors in an
orthonormal frame in any spacetime with a regular horizon pick up their
values from a horizon: $\lim_{r\rightarrow r_{+}}T_{\mu }^{\nu }(r)=T_{\mu
}^{\nu }(r_{+})$. Then the regularity condition on a horizon $%
T_{0}^{0}-T_{1}^{1}=0$ holds for a whole manifold. With the condition $%
r=r_{+}$, this means that both terms in entropy (\ref{S}), (\ref{2S})
(either ''normal'' or ''anomalous'' ones) are equal to zero, so $S_{q}=0$.
In particular, the previous result \cite{ent} for the Bertotti-Robinson
spacetime (for which $q=\sinh zy$ and $T_{\mu }^{\nu }\sim \delta _{\mu
}^{\nu }r_{+}^{-4}$\cite{kof}, \cite{paul}) is reproduced.

It is also worth paying attention to the degenerate case corresponding to
the ''ultraextreme'' case with $U^{\prime \prime }(r_{+})=V^{\prime \prime
}(r_{+})=0$. Then the limiting procedure elaborated in \cite{zasl2} leads to
the metric which is a direct product of two-dimensional Rindler space and a
sphere: 
\begin{equation}
ds^{2}=d\tau ^{2}l^{2}+dl^{2}+r_{+}^{2}d\omega ^{2}\text{.}  \label{rin2}
\end{equation}
In spite of $T_{\mu }^{\nu }\neq 0$ due to effects of curvature of
spacetime, the entropy of radiation $S_{q}=0$ according to the general
properties $T_{1}^{1}-T_{0}^{0}=0$ and $r=r_{+}$.

One reservation is in order here. The possibility to replace components $%
T_{\mu }^{\nu }$ by their horizon values due to taking the limit under
discussion implies that a horizon itself is regular in the original metric
in the extreme state. This is not the case for dilatonic black holes and, as
a result, the coefficient at the angular part does not turn into a constant
in this limit but retains some dependence on $l$ \cite{dil}.
Correspondingly, some residual dependence on $l$ may survive for $T_{\mu
}^{\nu }$ and there is no reason to expect $S_{q}=0$ in dilatonic
backgrounds. We will not, however, discuss this case here further.

Thus, for a generic metric obtained as a finite-temperature extreme limit of
nonextreme black holes with a regular extreme state the entropy of quantum
massless Hawking radiation $S_{q}=0$. Thus, the total entropy of such
systems dressed by Hawking radiation (which turn into spacetimes with
acceleration horizons in the limit under consideration) is equal to those of
bare ones.

\section{unruh effect and limits of spacetimes}

What physical explanation can be suggested for the property $S_{q}=0$?
Metrics discussed above share the following feature: they are obtained by a
special kind of the limiting transition $r\rightarrow r_{+}$. As a result,
spacetime picks up the sharp strip of near horizon geometry which expands
into a whole manifold with a finite Euclidean four-volume, so this is not
approximation but the example of taking the ''spacetime limit'' procedure
which maps an original manifold onto a new one \cite{ger}. In so doing, a
new spacetime inherits properties of a vicinity of a horizon where a metric
looks like that perceived by an accelerated observer, so as a matter of fact
we deal with the Unruh effect \cite{bir}. This effect, however, has a pure
kinematic nature and does not need the existence of a true black horizon; in
a sense, it is too week to gain nonzero entropy of Hawking radiation as
there are no ''true'' quanta of it. (To avoid possible confusion in
terminology, let me stress that we distinguish here the Unruh and Hawking
effects as connected with the presence of acceleration and true black hole
horizons, correspondingly. Thus, we use these terms in a way different from
the book \cite{wald}, where the thermal properties of the Hartle-Hawking
state are prescribed, by definition, to the Unruh effect in a curved
spacetime independently of the nature of a horizon --- in particular, in the
Schwarzschild background, i.e. in a true black hole metric. On the other
hand, the term ''Hawking effect'' is used therein to describe a dynamical
process of particles creation.)

Such a role of an acceleration horizon could serve as one more manifestation
of the kinematical nature of Hawking radiation which may or may not be
connected with an entropy associated with a horizon \cite{viss}. However, it
is worth noting that in our case, in contrast to what was discussed in \cite
{viss}, the zero-loop entropy connected with the information loss does exist
and kinematical properties of a spacetime reveal themselves only in
cancellation of the one-loop part of entropy.

In general, metric obtained after the limiting transition in question, is
curved; if quantum backreaction is neglected, it is the BR spacetime which
is nothing else than a direct product of anti-de Sitter space and a sphere $%
(AdS_{2}\times S_{2})$ since the curvature of $(\tau ,r)$ submanifold is a
negative constant. Such a spacetime has three independent Killing vectors 
\cite{lap} and an observer moving along a Killing orbit may feel horizons
with nonzero or zero Hawking temperature or see the absence of a horizon at
all that justifies purely kinematic nature of the effect in question.

In this sense there is some analogy in relationship between different
sections of BR spacetime, on one hand, and relationship between the
Minkowski and Rindler spaces, on the other one. Now we show that in the
properly adjusted ''large mass limit'' $r_{+}\rightarrow \infty $ this
analogy turns into literal correspondence. Let us write down the Lorentzian
form of the BR metric relevant for description of the extreme limit of
nonextreme black holes (BR1): 
\begin{equation}
ds^{2}=r_{+}^{2}(-dt^{2}\sinh ^{2}x+dx^{2}+d\theta ^{2}+d\phi ^{2}\sin
^{2}\theta )\text{.}  \label{br}
\end{equation}

This metric possesses a horizon at $x=0$ which, however, is not of black
hole type but rather is an acceleration horizon. One can perform the
transformation into another frame in which an observer moving along orbits
of another Killing vector will see no horizons at all \cite{lap}. Namely,
after the transformation 
\begin{equation}
\cosh t\sinh x=\sinh \chi \text{, }\cosh x=\cos \tilde{t}\cosh \chi
\label{tr}
\end{equation}
we arrive at the metric BR2 
\begin{equation}
ds^{2}=r_{+}^{2}(-d\tilde{t}^{2}\cosh ^{2}\chi +d\chi ^{2}+d\theta
^{2}+d\phi ^{2}\sin ^{2}\theta )\text{.}  \label{2br}
\end{equation}
Let us perform the transformation $\theta =\vartheta +\pi /2$, $x=l/r_{+}$, $%
\tilde{t}=T/r_{+}$, $\chi =Z/r_{+}$, $\vartheta =X/r_{+}$, $\phi =Y/r_{+}$.
Then after the limit $r_{+}\rightarrow \infty $ is taken, the metric (\ref
{br}) turns into the Rindler one 
\begin{equation}
ds^{2}=-dt^{2}l^{2}+dl^{2}+dX^{2}+dY^{2}\text{,}  \label{4rin}
\end{equation}
while (\ref{2br}) becomes the Minkowski metric 
\begin{equation}
ds^{2}=-dT^{2}+dZ^{2}+dX^{2}+dY^{2}\text{.}  \label{mink}
\end{equation}
Expanding eq. (\ref{tr}) in powers of $r_{+}^{-1}$ and retaining main
non-vanishing terms, we obtain the formulae 
\begin{equation}
Z=l\cosh t\text{, }Z^{2}-T^{2}=l^{2}  \label{zt}
\end{equation}
which describe just the connection between the Minkowski and Rindler
metrics. As far as the entropy of thermal gas is concerned, it remains zero
in the process of the limiting transition for both sections of the BR
spacetime (\ref{br}), (\ref{2br}). Thus, the analogy between the
Rindler/Minkowski and BR1/BR2 metrics in what concerns pure kinematic nature
of Hawking radiation and the property $S_{q}=0$ becomes literal in the limit
at hand.

It is worthwhile to note that, strictly speaking, the result $S_{q}=0$ for
the Rindler spacetime (\ref{4rin}) does not follow from the general formula (%
\ref{2S}) directly since the metric (\ref{4rin}) does not belong (in
contrast to (\ref{rin2})) to the limiting class of metrics (\ref{genmet})
for which (\ref{2S}) was derived. However, the above limiting relation makes
this property transparent since $S_{q}($Rindler$)=\lim S_{q}($BR$)=0$.
Nevertheless, it is also instructive to trace another kind of limiting
transition --- directly from the Schwarzschild metric since, as we will see,
it exhibits features similar to those for the limiting transition which
brings a nonextreme Reissner-Nordstr\"{o}m black hole to the extreme limit.
Consider the Schwarzschild metric 
\begin{equation}
ds^{2}=-dt^{2}(1-\frac{r_{+}}{r})+dr^{2}(1-\frac{r_{+}}{r}%
)^{-1}+r^{2}(d\theta ^{2}+\sin ^{2}\theta d\phi ^{2})\text{,}  \label{Sch}
\end{equation}
where -$\pi \leq \phi \leq \pi $, $0\leq \theta \leq \pi $. Here $r_{+}=2M$
is a horizon radius of a black hole with mass $M$. We will assume that a
black hole is situated in a cavity whose boundary ensures thermal
equilibrium between a black hole itself and its Hawking radiation (the
Hartle-Hawking state). Then, as was shown in \cite{rad}, the entropy of
massless Hawking radiation is described by eq. (\ref{ent1}) with $\left|
f^{\prime }(1)\right| =1$. Due to scale properties of massless quantum
radiation (\ref{tensor}) this equation can be rewritten as 
\begin{equation}
S_{q}=16\pi ^{2}\int_{w}^{1}duu^{-4}[f_{r}^{r}-f_{0}^{0}-f_{\mu }^{\mu }\ln
(u/w)]\text{,}  \label{ent}
\end{equation}
where $w=r_{+}/r$. The formulas (\ref{ent1}), (\ref{ent}) hold for any kind
of massless radiation including either bosons with generic type of coupling
to gravity or fermions \cite{mat}.

In the canonical ensemble the event horizon radius $r_{+}$ is not arbitrary
but is a function of either $r_{B}$ or a local Tolman temperature $\beta
^{-1}$on a boundary according to the equation \cite{york86} 
\begin{equation}
\beta =4\pi r_{+}\sqrt{1-\frac{r_{+}}{r_{B}}}\text{.}  \label{tol}
\end{equation}
Accounting for a finite size of a system has the crucial consequences for
thermodynamics: it allows one to define the canonical ensemble for black
holes in a self-consistent way, leads to the appearance of the stable branch
of solutions, etc. \cite{york86}. Now we will consider the large mass limit
which takes into account properly that the system has a finite size, so $%
r_{+}\leq r\leq r_{B}$. We will also assume that the limiting transition
preserves the value of $\beta $. These assumptions mean that the process of
limiting transition is performed in such a way that $r_{+}\rightarrow \infty 
$, $r_{B}\rightarrow \infty $, while a square root in eq. (\ref{tol}) tends
to zero and $r_{+}/r\rightarrow 1$ for all points of the manifold. Thus, the
coordinate $r$ becomes degenerate and is to be properly rescaled. In a
similar way, as the inverse Hawking temperature $T_{H}^{-1}=4\pi
r_{+}\rightarrow \infty $, the time coordinate needs to be rescaled too.
Before taking such a limit, let us perform the change of variable $\theta
=\vartheta +\pi /2$, so $-\pi /2\leq \vartheta \leq \pi /2$ and introduce
new coordinates according to 
\begin{equation}
\vartheta =X/r_{+}\text{, }\phi =Y/r_{+}\text{, }t=\tau /2\pi T_{H}=2\tau
r_{+}\text{, }r=r_{+}+l^{2}/4r_{+}\text{.}  \label{new}
\end{equation}
Then after the limiting transition at hand the original metric (\ref{Sch})
turns into the Rindler metric (\ref{4rin}). It follows from (\ref{tensor})
that $T_{\mu }^{\nu }\rightarrow 0$ in accordance with the fact that in the
state of thermal equilibrium the stress-energy tensor in the Hartle -
Hawking state cancels for the Rindler metric \cite{ginz}. It is worth noting
that in the Hartle - Hawking state, the components $T_{\mu }^{\nu }$ of the
stress-energy tensor in an orthonormal frame are finite on the event horizon
of the Schwarzschild black hole, so the quantities $f_{\mu }^{\nu }$ are
also finite at $u\rightarrow 1$. Therefore, it is seen from (\ref{ent}) that 
$S_{q}\rightarrow 0$ in the limit at hand: the thermodynamical entropy of
Hawking radiation cancels for a Rindler wedge. In a sense, thermal gas of
Rindler quanta is a rather peculiar object from the thermodynamic point of
view: in spite of its temperature being nonzero, either its energy defined
as $-\int d^{3}x\sqrt{g}T_{0}^{0}$ or the entropy are equal to zero. I
stress that the statement $S_{q}=0$ is related just to the thermodynamic
entropy (i.e. the quantity just having direct physical meaning) and should
not be confused with the properties of statistical-mechanical one \cite
{solod}, \cite{zerb} (in a similar way, the property $S_{q}=0$ for a thermal
gas in the BR background \cite{ent} should not be confused with the behavior
of quantum correction to the entropy of a black hole itself \cite{man}).

It is worth stressing that the limiting procedure for any spacetime is not
unique and depends, for example, on a particular choice of a coordinate
system in which parameters of a system tend to their limiting values \cite
{ger}. In our case the limit $r_{+}\rightarrow \infty $ is distinguished by
the demand that a local temperature on a boundary is fixed, so it has clear
thermodynamic meaning. In so doing, however, the boundary itself drastically
changes: a sphere turns into a plane. As in the limit at hand $%
r_{+}\rightarrow \infty $ in such a way that $r_{+}/r_{B}\rightarrow 1$, the
zero-loop entropy of a black hole, equal to its Bekenstein-Hawking value,
behaves like $S_{0}=\pi r_{+}^{2}\simeq \pi r_{B}^{2}\simeq A/4$ where $%
A=\int dXdY$ is the surface area of a plane $l=const$, so $S_{0}$ diverges
but the entropy per unit area is finite \cite{lf}. It is worthwhile to note
that the derivation of the formula for the entropy $S_{0}$ in \cite{lf}
relies at once on the metric of the Rindler wedge and the term $A/4$ comes
from a boundary, so the connection between this term and the horizon
remained not quite clear (any surface $l=const$ has the infinite area $A$
for the Rindler metric). Meanwhile, the limiting transition performed above
clearly shows that the Rindler wedge inherits the formula $S_{0}=A/4$ from
the Schwarzschild spacetime where it originates from an event horizon.

It is interesting that, although $A\rightarrow \infty $, the proper distance 
$L$ between a horizon and any other fixed point outside, including a
boundary, is finite. Indeed, in the Schwarzschild metric we have $L=r_{+}[%
\sqrt{x(x-1)}+\ln (\sqrt{x}+\sqrt{x-1})]$ where $x=r/r_{+}$. In the limit $%
r_{+}\rightarrow \infty $ performed in coordinates (\ref{new}), $%
L\rightarrow l$. In other words, we have a plane situated at the proper
distance $l_{B}$ from the origin of coordinates, where $l_{B}$ is the value
of the $l$-coordinate of the boundary in the original Schwarzschild metric,
this plane having the same temperature as a boundary sphere in the
Scwharzschild metric.

In general, the total entropy of a system possessing a horizons comes,
according to (\ref{total}), from either the horizon (the term $S_{0}$)
itself or quantum fields (the term $S_{q}$). The fact that in our case the
entropy is determined by the area of the horizon only, this horizon having
kinematical nature (the acceleration horizon instead of the black hole one),
manifests the kinematic character of the Unruh effect in the given context.

Recently, an interesting interpretation of the Hawking effect as the Unruh
one in some embedding auxiliary flat space of higher dimensionality has been
suggested \cite{des}. On the contrary, in our approach we trace the passage
from one spacetime to another remaining within a physical four-dimensional
curved manifold. And what we want to stress is that the equivalence between
two effects traced in \cite{des} for temperatures and zero-loop entropies,
breaks down for the entropies of quantum radiation.

\section{general first law for acceleration horizons}

The Euclidean canonical action for a spherically-symmetrical bounded
self-gravitating charged system obeying the Hamiltonian constraint and Gauss
law reads \cite{braden} 
\begin{equation}
I=\beta E-S-\beta \phi e\text{,}  \label{action}
\end{equation}
where $\beta $ is an inverse Tolman temperature on a boundary, $\phi $ is a
blueshifted potential difference between the horizon and boundary, and $e$
is charge. For a given set of boundary data ($\beta $, $r_{B}$, $\phi $) a
small variation in a horizon radius gives, according to the action principle 
$\delta I=0$, the form of the first law under conditions that all the field
equations are satisfied, so terms arising due to equations of motions cancel 
\cite{quasi}. The energy $E=4\pi r_{B}^{2}\varepsilon $, where quasilocal
energy density $\varepsilon $ entering a thermodynamic energy $E$ is equal
to ($k-k_{0}$)$/8\pi $ \cite{canon}, \cite{quasi}, $k$ is an extrinsic
curvature of two-dimensional boundary embedded into a three-dimensional
space, $k_{0}$ is that for the same boundary metric embedded into a
reference flat space to have $E=0=I$ for a flat metric. In a
spherically-symmetrical spacetime of the form (\ref{spher}) $E=r_{B}[1-\sqrt{%
V(r_{B})}]$, where the first contribution corresponds to the flat space term 
$k_{0}$.

In attempting to apply the first law to metrics under discussion which have $%
r=cons=r_{+}$ one immediately faces the following oddities. The term with $k$
in energy is equal to zero identically. It follows either from definition of 
$k$ or from the above formula for $E$. Indeed, the metric in question is
obtained by the limit $r\rightarrow r_{+}$ for all points of manifold
including a boundary. As a result, the coefficient $V$ in the formula for $E$
picks up its value from a horizon where $V=0$. Apart from this, the quantity 
$r_{+}$ cannot be any longer considered as a free parameter independent of
boundary data. Now the two-dimensional metric induced on a horizon coincides
with that of a boundary. These two circumstances do not mean, however, that
the first law loses its sense. Rather, it leads us to the necessity to
consider its extended form including from the very beginning the
contribution from the changes of a boundary metrics. According to \cite
{quasi}, \cite{action}, such a contribution can be represented as $-\frac{1}{%
2}\int \delta \sigma _{ab}s^{ab}\sqrt{\sigma }\beta $ where indices $a$,$b$
are related to a two-dimensional boundary with the metric $\sigma _{ab}$, $%
s^{ab}$ are components of spatial stresses. For a spherically symmetrical
spacetime with $\beta =const$ on a boundary this reduces to $\lambda \delta
r_{B}$ where $\lambda =-4\pi r_{B}\beta s_{a}^{a}$. It follows from eq.
(6.9) of \cite{quasi} that, in our notations, $\lambda =2\pi [b-\frac{%
\partial (br)}{\partial l}]_{B}.$ Apart from gravitational contribution, we
must take into account the change of the action of quantum fields $8\pi
^{2}\gamma \delta r_{B}$ with $\gamma $ from eq. (\ref{gamma}). Comparing
with (\ref{action}), we have 
\begin{equation}
\beta \delta E-\beta \phi \delta e-\delta S=2\pi \{[b-\frac{\partial (br)}{%
\partial l}]_{B}+4\pi r_{B}^{-1}\int_{0}^{1}dy\sqrt{\tilde{g}}T_{\mu }^{\mu
}\}\delta r_{B}\text{.}  \label{first}
\end{equation}
In such a form the first law must hold for any spacetime of the form (\ref
{genmet}) under the presence of an electromagnetic field. Now we apply it to
the spacetimes which are the $r\rightarrow r_{+}$ limits of (\ref{spher}) in
the sense discussed above.

First, consider the classical BR spacetime for which in (\ref{genmet}) $%
b=r_{+}\sinh l/r_{+}$, $T_{\mu }^{\nu }$ is neglected. The energy $%
E=r_{B}=r_{+}$. Integrating the Maxwell equation $F_{;\mu }^{0\mu }\equiv
(F^{01}\sqrt{g})_{^{\prime }}/\sqrt{g}=0$ it is easy to find the value of $%
\phi \equiv b_{B}^{-1}[A_{0}(1)-A_{0}(0)]$, where $A_{0}(1)-A_{0}(0)$ is the
difference of electrostatic potentials $A_{0}$ between a horizon and
boundary: $\phi =\tanh l/2r_{+}$ where we have taken into account that for a
BR spacetime the charge $e=r_{+}$ (see for details below). Substituting $%
\beta =2\pi b_{B}$ and $S=\pi r_{+}^{2}$ into (\ref{first}) we see that the
first law is satisfied.

Let now quantum backreaction be taken into account, the total stress-energy
tensor $T_{\mu }^{\nu (tot)}=T_{\mu }^{\nu (em)}+T_{\mu }^{\nu }$
representing the sum of contributions from an electromagnetic field and
quantum one. For a metric (\ref{genmet}) with $r=r_{+}=const$ the
nonvanishing components of the Einstein tensor are $%
G_{0}^{0}=-1/r_{+}^{2}=G_{1}^{1}$, $G_{2}^{2}=G_{3}^{3}=b^{-1}\frac{\partial
^{2}b}{\partial l^{2}}$. If the Gauss law $\frac{\partial A_{0}}{\partial l}%
=eb/r_{+}^{2}$ is taken into account, the electromagnetic part of the
energy-momentum tensor has the standard form $T_{\mu }^{\nu (em)}=e^{2}/8\pi
r_{+}^{4}diag(-1,-1,1,1)$. For a massless radiation the stress-energy in the
BR background is 
\begin{equation}
T_{\mu }^{\nu }=\frac{B}{8\pi }\delta _{\mu }^{\nu }r_{+}^{-4}  \label{tenbr}
\end{equation}
\cite{kof}, \cite{paul}, where $B=const$ and the factor $(8\pi )^{-1}$ is
introduced for convenience. Einstein equations give us $e^{2}=r_{+}^{2}+B$, $%
b=\rho \sinh \rho ^{-1}l$ where $\rho ^{-2}=r_{+}^{-2}(1+2B/r_{+}^{2})$. In
the main approximation with respect to $B$ we have $\gamma =\frac{B}{2\pi
r_{+}}(\cosh l_{B}/r_{+}-1)$. From the Gauss law it follows that $\phi
\equiv b^{-1}[A_{0}-A_{0}(0)]=\frac{e}{r_{+}^{2}\rho }\tanh \frac{\rho l}{2}$%
. Substituting these expressions into (\ref{first}), we may check directly
that the general first law is satisfied provided $\delta S=2\pi r_{+}\delta
r_{+}$. Integrating this equality we obtain that $S=\pi r_{+}^{2}+c$ where $%
c $ is some constant. Here the first term represent Bekenstein - Hawking
entropy whereas the second one is responsible for Hawking radiation. From
the demand that the second contribution vanishes when a boundary approaches
the surface of a horizon (no room for radiation) we obtain that $c=0$. Thus,
we arrive at the same conclusion as was made above: entropy of Hawking
radiation in the BR background is equal to zero exactly.

\section{quantum-corrected geometry of spacetimes with acceleration horizons}

Limiting geometries found in \cite{zasl2} relied on the general assumptions
of the limiting transition from nonextreme black hole metrics to the extreme
state with a finite local temperature in any point between a horizon and
physical boundary. No field equations with or without backreaction of
quantum fields on a metric were used in \cite{zasl2}. Meanwhile, account for
such equations restricts strongly the possible type of limiting metrics. As
follows from the formulae of the previous section, the quantum-corrected BR
spacetime has the form 
\begin{eqnarray}
ds^{2} &=&d\tau ^{2}\rho ^{2}\sinh ^{2}\frac{l}{\rho }+dl^{2}+r_{+}^{2}d%
\omega ^{2}\text{, }\phi =\frac{e}{r_{+}^{2}\rho }\tanh \frac{\rho l}{2}%
\text{,}  \label{cor} \\
\rho ^{-2} &=&r_{+}^{-2}(1+\frac{2B}{r_{+}^{2}})\text{, }e^{2}=r_{+}^{2}+B%
\text{.}  \nonumber
\end{eqnarray}
There is also another solution 
\begin{equation}
ds^{2}=d\tau ^{2}\rho ^{2}e^{2\rho l}+dl^{2}+r_{+}^{2}d\omega ^{2}
\label{ext}
\end{equation}
with the same $\rho $.

Recently it was argued by Solodukhin that the product spacetime $%
AdS_{2}\times S_{2}$ is an exact solutions of semiclassical field equations
with quantum backreaction taken into account \cite{ads}. In fact, as in $(r$,%
$\tau )$ submanifold the curvature $R_{2}=-2\rho ^{-2}$ is constant and $%
R_{2}<0$, our formula (\ref{cor}) is in conformity with this statement. This
metric is based on the perturbative expression for $T_{\mu }^{\nu }$ valid
in the region $\left| B\right| \ll r_{+}^{2}$. It is instructive, however,
to extend (not quite rigorously) a semiclassical approach to pure quantum
domain for which $\left| B\right| \sim r_{+}^{2}$. Then for $B=-\left|
B\right| $ the $(_{2}^{2})$ equation gives us a qualitatively new type of
solutions if $\left| B\right| \geq r_{+}^{2}/2$: 
\begin{eqnarray}
ds^{2} &=&d\tau ^{2}\sigma ^{2}\sin ^{2}\frac{l}{\sigma }+dl^{2}+r_{+}^{2}d%
\omega ^{2}\text{, }\phi =\frac{e\sigma }{r_{+}^{2}}\tan \frac{l}{2\sigma }%
\text{,}  \label{sin} \\
\sigma ^{-2} &=&r_{+}^{-2}(\frac{2\left| B\right| }{r_{+}^{2}}-1)\text{, }%
e^{2}=r_{+}^{2}-\left| B\right| \text{.}  \nonumber
\end{eqnarray}
Formally, the metric (\ref{sin}) is obtained from (\ref{cor}) by the
substitution $\rho =i\sigma $. It is seen from (\ref{sin}) that the solution
of the form (\ref{sin}) may exist only under the condition 
\begin{equation}
r_{+}^{2}/2\leq \left| B\right| \leq r_{+}^{2}  \label{ineq}
\end{equation}
, i.e. for strong backreaction, and in this sense is pure quantum, the
curvature of $(r$,$\tau )$ submanifold $R_{2}=2\sigma ^{-2}=const\geq 0$. In
these respects the $(r$,$\tau )$ part of (\ref{sin}) resembles the one found
in \cite{2d} for 2D dilaton gravity. Thus, in addition to the $AdS_{2}\times
S_{2}$ found in \cite{ads}, in our problem there exists also spacetime which
is a direct product of de Sitter space and a sphere $(dS_{2}\times S_{2})$.
As for this solution $T_{\mu }^{\nu (em)}=0$ and $B<0$, the energy density
is positive everywhere including a horizon. In this respect it can be
considered as a BR-like counterpart of black holes which may possess the
extreme state due to positive energy density on a horizon whose existence
was qualitatively conjectured in \cite{balb}.

For the solution under discussion the ratio of squared radii of two pieces
of spacetime $0\leq r_{+}^{2}/\sigma ^{2}\leq 1$. The minimum value of this
ration is achieved at $R_{2}=2\sigma ^{-2}=0$, $\left| B\right| =r_{+}^{2}/2$
when we obtain the spacetime $Rindler_{2}\times S_{2}$ (\ref{rin2}) with the
electrostatic potential $\phi =2^{-3/2}l/r_{+}$. In contrast to (\ref{cor})
- (\ref{sin}), the solution at hand has no 2D counterpart: in the latter
case \cite{2d} a metric can be flat only under condition that either an
electromagnetic field or backreaction cancel whereas now either each
contribution separately or their sum differs from zero: $%
T_{0}^{0(tot)}=T_{1}^{1(tot)}=-1/8\pi r_{+}^{2}$. Such a spacetime can be
regarded as the example of physical realization of the ultraextreme limit of
nonextreme black holes \cite{zasl2} that shows how a Rindler metric may
appear as a nontrivial result of special tuning between electromagnetic
forces and quantum backreaction: $e^{2}=\left| B\right| ^{2}$ and tangential
stresses vanish, $T_{2}^{2(tot)}=T_{3}^{3(tot)}=0$.

The maximum value of $r_{+}^{2}/\sigma ^{2}$ corresponds to $e=0$, $%
B=-r_{+}^{2}$, $\sigma =r_{+}$. The possibility $e=0$ due to quantum effects
was suggested by Solodukhin \cite{ads} for the $AdS_{2}\times S_{2}$
solution. Our formulae, however, do not admit such a possibility for the $%
AdS_{2}\times S_{2}$ case since it is inconsistent with the property $%
R_{2}<0 $ according to (\ref{cor}). This difference in properties of
solutions under discussion can be explained by the fact that we are dealing
with a conformally invariant scalar field, whereas in \cite{ads} this field
has minimal coupling. Meanwhile, for $dS_{2}\times S_{2}$ solutions $e=0$ is
indeed possible, in which case radii of both two-dimensional subspaces
coincide.

It is instructive to suggest qualitative explanation of rather unusual
consequences in the structure of spacetime caused by quantum backreaction.
The key moment consists in the structure of the energy-momentum tensor (\ref
{tenbr}). It is seen from eq. (\ref{tenbr}) that the stress-energy tensor in
the BR metric mimics the effect of the cosmological constant: $T_{\mu }^{\nu
}=\Lambda _{eff}\delta _{\mu }^{\nu }$, where the effective cosmological
constant $\Lambda _{eff}=-Br_{+}^{-4}$. This constant is absent on the
classical level and is caused in our case by quantum effects entirely. If $%
B<0$, $\Lambda _{eff}>0$. If a system possesses either an electric charge or
the positive cosmological constant, we have, in general, the
Reissner-Nordstrom-de Sitter solution (RNdS) with three horizons - the inner
one $r_{i}$, the outer black hole horizon $r_{o}$ and the cosmological one $%
r_{c}$. In the particular case, when the radii $r_{o}$ and $r_{c}$ merge,
one obtains the charged version of the Nariai solution \cite{nar}. It can be
obtained as the so-called cold limit of the RNdS metric \cite{romans}. In
this limit, the volume in the region $r_{o}<r<r_{c}$ remains finite despite $%
r_{o}\rightarrow r_{c}$, and the new metric, arising as a result of the
limiting transition, looks just like (\ref{sin}). The surface gravity of
each horizon in question tends to zero (that motivates the name ''cold'')
but it is essential that the physical Tolman temperature in every point
outside the horizon remains finite nonzero quantity (in Ref. \cite{dist}
such a kind of limiting transitions is considered in a general setting
without specifying the concrete type of the metric). Substituting $\left|
B\right| =\Lambda _{eff}r_{+}^{4}$ into eq. (\ref{ineq}), we obtain
inequalities inherent to the charged Nariai solution, $e^{2}<\Lambda
_{eff}r_{+}^{4}$ and $\Lambda _{eff}r_{+}^{2}<1$ (the modern discussion of
the Nariai solution and its properties as well as a number of references can
be found in \cite{bousso}). The case of the equalities corresponds to the
ultracold one (see below). The physical interest in spacetimes under
discussion is dictated, in particular, by their role in pair creation of
black holes in cosmological backgrounds (see, for instance, Refs. \cite
{mann95}, \cite{cosm} and references therein).

In the case when all three horizons merge ($r_{i}\rightarrow
r_{o}\rightarrow r_{c}$) the so-called ultracold limit of the RNdS
spacetimes arises \cite{romans}. The corresponding Euclidean metric reads (%
\ref{rin2}) \cite{mann95} and, thus, coincides with the quantum-corrected
geometry obtained by us above.

Thus, the structure of the resulting spacetime depends on the sign of $%
\Lambda _{eff}$ and can be thought of as the result of different types of
limiting transitions to the extreme state. If $\Lambda _{eff}<0$, we have $%
AdS_{2}\times S_{2}$ which can be considered as a result of the extreme
limit for nonextreme black holes \cite{zasl1}, \cite{zasl2}. If $\Lambda
_{eff}>0$, the metric has the form $dS_{2}\times S_{2}$ (\ref{sin}) which
appears after the limiting transition to the state with $r_{o}=r_{c}\neq
r_{i}$, or $Rindler_{2}\times S_{2}$ when radii of all three horizons merge.

Let me stress that the qualittaively new point as compared with Refs. \cite
{romans}, \cite{bousso}, \cite{mann95}, \cite{cosm} and references therein
consists in that the $\Lambda $-term in our problem is absent classically,
so the appearance of analogs of the limiting forms of the RNdS solutions is
a pure quantum effect. It is also worth noting that we did not consider the
RNdS with the consequent limiting transition but started at once with the
BR-like spacetimes and showed how account for backreaction of quantum fields
may change the properties of a spacetime.

The constant $B$, responsible for quantum effects, can be written as $%
al_{0}^{2}$, where $l_{0}$ is the Planck length, $a$ is numerical
coefficient. According to (\ref{cor}) - (\ref{sin}), the solution with $e=0$
is possible only for $a=-\left| a\right| <0$. In particular, a massless
conformally invariant field for which $a=(2880\pi ^{2})^{-1}>0$ \cite{paul}
seems to be not a suitable candidate for such type of solutions. If $e=0$,
the radius $r_{+}$ which measures the curvature of a sphere acquires
planckian scale in agreement with Solodukhin's observation: $r_{+}=\left|
a\right| l_{0}$ (In fact, even $r_{+}\ll l_{0}$, but one can attain $%
r_{+}\sim l_{0}$ for a sufficiently large number of field species).

One reservation is in order. From the formal viewpoint, the metric (\ref{cor}%
) was based on a semiclassical expression for the stress-energy tensor $%
T_{\mu }^{\nu }=(8\pi )^{-1}Br_{+}^{-4}\delta _{\mu }^{\nu }$ calculated on
a given BR background, so the extension to the domain $\left| B\right| \sim
r_{+}^{2}$ is nothing else than extrapolation. Nevertheless, the striking
similarity to the 2D case where a semiclassical stress-energy tensor is
known exactly, strongly supports the validity of found solutions. More
rigorous justification of possibilities indicated in \cite{ads} and in the
present paper needs a fully self-consistent quantum treatment. One cannot
exclude in advance the existence of pure quantum solutions with $e=0$ for
any kind of quantum field.

Anyway, the indicated class of solutions (\ref{sin}) indebted entirely to
quantum effects hints that strong backreaction can modify the structure of
spacetime qualitatively and, in particular, change the sign of curvature.
Drastic changes may also happen not only to solutions (\ref{cor})-(\ref{sin}%
) themselves but also with black holes whose near-horizon geometry may be
approximated by these constant curvature solutions. One such possibility
consists in the existence of an extreme quantum Schwarzschild-like black
hole with a zero surface gravity \cite{balb}, \cite{ads}. Here we would like
to draw attention to another possibility: the appearance, according to (\ref
{sin}), of the solution with a cosmological horizon.

Thus, on one hand, quantum effects preserve the general form of the metric
as a direct product of two two-dimensional submanifolds in accordance with 
\cite{ads}. On the other hand, the concrete set of possible kinds of such a
structure includes not only spacetimes with $R_{2}<0$ indicated in \cite{ads}
but also those with both other possible types $R_{2}>0$ and $R_{2}=0$.

\section{conclusion}

In general, one may distinguish three areas of application of thermodynamic
approach to systems with horizons: nonextreme black holes, extreme black
holes and acceleration horizons. The present paper is devoted to the third
case and, in this sense, fills a gap in the relationship between
thermodynamics and horizon mechanics. A typical representative of the
spacetimes with an acceleration horizon is the BR metric. The interest to
different aspects of BR-like spacetimes has increased in recent years \cite
{recent}. Here, we considered quantum-corrected BR-like spacetimes and
showed that their acceleration horizons exhibit the following universal
property: the entropy of Hawking massless radiation $S_{q}=0$. The procedure
of taking spacetimes limits, with the help of which the metrics with an
acceleration horizons are obtained from black hole ones, showed that the
result $S_{q}=0$ is intimately connected with the Unruh effect rather than
with the Hawking one. The general first law does have sense for the metrics
in question in spite of some restrictions on the variation procedure which
now implies that the boundary radius is to be varied together with that of
the horizon. If formulated properly, this law confirms that the entropy of
radiation does not contribute to thermodynamics of a system. The result $%
S_{q}=0$ can serve as a test for checking various renormalization schemes in
calculations of quantum entropy of black holes which possess the extreme
state.

While quantum effects do not reveal themselves directly in thermodynamics of
acceleration horizons, they can have crucial consequences for a structure of
spacetime. In particular, the strong backreaction seems to lead to the
possible change of the sign of two-dimensional $R_{2}$ curvature of $(r$,$%
\tau )$ submanifold and the appearance of the quantum version of the Nariai
solution with cosmological horizon but without the cosmological constant or
to the possibility to have a flat $(r$,$\tau )$ submanifold as an exact
solution of {\it semiclassical} field equations.

In the present paper we restricted ourselves by the case of massless fields
since their scale properties simplify the problem at once in two points: (i)
the action depends on one variable and (ii) the thermal stress-energy tensor
has the general structure (\ref{tensor}) (see Sec.II above). Meanwhile, it
is of interest to obtain the formula for the entropy of quantum massive
fields in terms of the stress-energy tensor which would replace eq. (\ref{2S}%
) derived for massless ones, and check the validity of the property $S_{q}=0$%
. The case of massive fields is especially important in connection with the
issue of quantum renormalization of the black hole entropy. It was shown in 
\cite{dem} without using the brick-wall model that Pauli-Villars
regularization correctly reproduces the Bekenstein-Hawking entropy, if the
gravitational constant is properly renormalized (this approach was
elaborated further for many-dimensional cases in \cite{kim}). Therefore, the
desired formula for the thermodynamic entropy, being combined with the
results of \cite{dem}, would enable us, in particular, to trace in detail
the behavior of different parts of the black entropy near the extreme state.
We hope to address this issue in a subsequent research.

Apart from extending the approach to another kinds of fields, it is also
worthwhile to consider more general geometries - in particular, spacetimes
which may be obtained by the spacetime limits taken in the background of
distorted black holes \cite{dist}. Of special interest is the problem of
finding either quantum geometries or the stress-energy tensor in a
background with acceleration horizons in a fully self-consistent manner.

In this paper, we restricted ourselves to the limiting states of type 2
which are obtained by certain limiting transitions to the extreme state
within the topological sector of nonextreme black holes. The separate issue
deserving special treatment is the influence of quantum backreaction on
states of type 1 which represent topologically true extreme black holes.

\section{acknowledgments}

I am grateful to Ted Jacobson for comment on my paper \cite{zasl2} with the
interpretation of limiting geometries found therein as describing the Unruh
effect in the $AdS_{2}$ background. I am also grateful to Sergey Solodukhin
for helpful correspondence. This work is supported by the International
Science Education Program, grant \# QSU082068.


% insert suggested PACS	numbers	in braces on next line

% body of paper	here

% now the references. delete or	change fake bibitem. delete next three
%   lines and directly read in your .bbl file if you use bibtex.

% Here is an example of	the general form of a figure:
% that you will	use with \ref{}	command	in the braces of the \label{} command.
%
% \begin{figure}
% \caption{}
% \label{}
% \end{figure}

% tables follow	here
%
% Here is an example of	the general form of a table:
% that you will	use with \ref{}	command	in the braces of the \label{} command.
% \begin{table}
% \caption{}
% \label{}
% \end{tabular}
% \end{table}

\end{document}